\documentstyle[12pt]{article}

\newlength{\extralineskip}

\def\inbar{\,\vrule height1.5ex width.4pt depth0pt}
\def\IC{\relax\hbox{$\inbar\kern-.3em{\rm C}$}}
\def\IR{\relax{\rm I\kern-.18em R}}
\def\IZ{\relax\ifmmode\mathchoice
{\hbox{\kern-.4em Z}}{\hbox{Z\kern-.4em Z}}
{\lower.9pt\hbox{Z\kern-.4em Z}}
{\lower1.2pt\hbox{Z\kern-.4em Z}}\else{Z\kern-.4em Z}\fi}
\def\pvint{{\int\!\!\!\!\!\!-}}

\def\bea{\begin{eqnarray}}
\def\eea{\end{eqnarray}}

\def\beq{\begin{equation}}
\def\eeq{\end{equation}}
\def\ba{\beq\new\begin{array}{c}}
\def\ea{\end{array}\eeq}

\def\tr{~{\rm tr}~}

\def\ps2{{\bar{\psi}\psi}}
\def\e{~{\rm e}}
\def\pvint{{\int\!\!\!\!\!\!-}}
\makeatletter
\newdimen\normalarrayskip              
\newdimen\minarrayskip                 
\normalarrayskip\baselineskip
\minarrayskip\jot
\newif\ifold             \oldtrue            \def\new{\oldfalse}
\def\arraymode{\ifold\relax\else\displaystyle\fi} 
\def\@arrayskip{\ifold\baselineskip\z@\lineskip\z@
     \else
     \baselineskip\minarrayskip\lineskip2\minarrayskip\fi}
\def\@arrayclassz{\ifcase \@lastchclass \@acolampacol \or
\@ampacol \or \or \or \@addamp \or
   \@acolampacol \or \@firstampfalse \@acol \fi
\edef\@preamble{\@preamble
  \ifcase \@chnum
     \hfil$\relax\arraymode\@sharp$\hfil
     \or $\relax\arraymode\@sharp$\hfil
     \or \hfil$\relax\arraymode\@sharp$\fi}}
\def\@array[#1]#2{\setbox\@arstrutbox=\hbox{\vrule
     height\arraystretch \ht\strutbox
     depth\arraystretch \dp\strutbox
     width\z@}\@mkpream{#2}\edef\@preamble{\halign \noexpand\@halignto
\bgroup \tabskip\z@ \@arstrut \@preamble \tabskip\z@ \cr}%
\let\@startpbox\@@startpbox \let\@endpbox\@@endpbox
  \if #1t\vtop \else \if#1b\vbox \else \vcenter \fi\fi
  \bgroup \let\par\relax
  \let\@sharp##\let\protect\relax
  \@arrayskip\@preamble}

\begin{document}
\begin{titlepage}
\rightline{\baselineskip=12pt\vbox{\halign{&#\hfil\cr
UBC/S-94/1&\cr
hep-th/9410214 &\cr
{   }&\cr \today&\cr}}}
\vspace{0.5in}
\begin{center}
{\Large\bf Critical Behaviour of a Fermionic Random Matrix Model at
Large-$N$}\\
\medskip
\vskip0.5in
\baselineskip=12pt

\normalsize {\bf Nicole Marshall}\footnote{Work supported in part by the
Natural Sciences and Engineering Research Council of Canada.}, {\bf Gordon W.
Semenoff} $^1$ and {\bf Richard J. Szabo}\footnote{Work supported in part
by a University of British Columbia Graduate Fellowship.}
\smallskip
\medskip

{\it Department of Physics\\ University of British Columbia\\
Vancouver, British Columbia, Canada V6T 1Z1}\smallskip
\end{center}
\vskip1.5in

\begin{abstract}
\baselineskip=12pt

We study the large-$N$ limit of adjoint fermion one-matrix models.  We
find one-cut solutions of the loop equations for the correlators of
these models and show that they exhibit third order phase transitions
associated with $m$-th order multi-critical points with string
susceptibility exponents $\gamma_{\rm str}=-1/m$. We also find
critical points which can be interpreted as points of first order
phase transitions, and we discuss the implications of this critical behaviour
for the topological expansion of these matrix models.

\end{abstract}
\end{titlepage}
\newpage

\baselineskip=18pt

Hermitian matrix models are the classic example of a $D=0$ quantum
field theory where 'tHooft's topological large-$N$ expansion
\cite{tH} can be solved explicitly \cite{bipz,biz}. They have
recently been of interest in the study of the statistical mechanics of
random surfaces \cite{adf} particularly for non-perturbative
approaches to lower dimensional string theory
\cite{bk}. There, the large-$N$ expansion coincides with the genus
expansion and the large-$N$ limit exhibits phase transitions which
correspond to the continuum limits of the discretized random surface
theories \cite{bk}. Unitary matrix models also play a role in
2-dimensional QCD \cite{gw}, mean-field computations in lattice gauge
theory \cite{di} and various other approaches to higher dimensional
gauge theories such as induced QCD \cite{km}.  In this Letter we study
a matrix model where the degrees of freedom are matrices whose
elements are anticommuting Grassmann numbers. The partition function
is
\begin{equation}
Z=\int d\psi~d\bar\psi~\e^{N^2\tr V(\bar\psi\psi)}
\label{part}
\end{equation}
where $V$ is a polynomial potential,
\beq
V(\bar\psi\psi)=\sum_{k=1}^K\frac{g_k}{k}(\bar\psi\psi)^k~~~~,
\label{pot}
\eeq
$\psi$ and $\bar\psi$ are independent $N\times N$ matrices with
anticommuting nilpotent elements and the integration measure,
$d\psi~d\bar\psi\equiv\prod_{i,j}d\psi_{ij}~d\bar\psi_{ij}$, is
defined using the usual rules for integrating Grassmann variables,
$\int d\psi_{ij}~\psi_{ij}=1$, $\int d\psi_{ij}~1=0$. We normalize
all traces here and in the following as $\tr
V\equiv\frac{1}{N}\sum_{i}V_{ii}$.

Matrix models of this kind have been studied recently by Makeenko and Zarembo
\cite{mz}, and Ambj\o rn, Kristjansen and Makeenko \cite{akm}.  They are
motivated by models of induced gauge theories using adjoint matter
where the Yang-Mills interactions of gluons are induced by loops with
heavy adjoint scalar fields \cite{km} or other kinds of matter such as
heavy adjoint fermions \cite{mz,khm}.  Makeenko and Zarembo
\cite{mz} have shown that the adjoint fermion matrix model
(\ref{part}) has many of the features of the more familiar Hermitian
one-matrix model in the large-$N$ limit \cite{bk,kaz},
including multi-critical behaviour with a third order phase transition
and string susceptibility with critical exponent $\gamma_{\rm
str}=-1/m$, $m\in\IZ^+$.  They also showed that the loop equations for
the model (\ref{part}) are identical to those for the Hermitian
one-matrix model with generalized Penner potential
\cite{akm,pen}
\begin{equation}
Z_P=\int\prod_{i,j}d\phi_{ij}~\e^{-N^2\tr\left( V(\phi)-2\log\phi\right)}
\label{penner}
\end{equation}
However, the loop equations for the two models should be solved with
different boundary conditions and the solution beyond the leading
order in large-$N$ is different in the two cases.  It was argued in
\cite{mz} and \cite{akm} that, as the Penner model corresponds to
a certain statistical theory of triangulated random surfaces, the
fermionic matrix model corresponds to a similar theory where the genus
expansion has alternating signs. The resulting convergence of the sum
over genera is reflected in the feature of the fermionic matrix model
(\ref{part}) that its partition function and observables are
well-defined since the integrals over Grassmann variables always
converge, in contrast to Hermitian matrix models
\cite{bk,kaz} where, typically, in the region of interest the
integration over Hermitian matrices diverges (reflecting of course the
divergence of the genus sum in the random surface model).

In fact, for any polynomial potential the integration over anticommuting
variables in (\ref{part}) can be formally performed by inserting the
matrix-valued delta function $1=\int\prod_{i,j}d\phi_{ij}~\delta(\phi-\bar
\psi\psi)$, where $\phi$ is a Hermitian matrix, and using the identity
\beq
\int\prod_{i,j}\frac{d\lambda_{ij}}{2\pi}~\int d\psi~d\bar\psi~\e^{i\tr\lambda
(\phi-\bar\psi\psi)}=\int\prod_{i,j}\frac{d\lambda_{ij}}{2\pi}~
{\det}^N(-i\lambda)\e^{i\tr\lambda\phi}={\det}^N\left(-\frac{\partial}
{\partial\phi}\right)\delta(\phi)
\eeq
to obtain
\begin{equation}
Z={\det}^N\left(\frac{\partial}{\partial\phi}\right)\e^{N^2\tr
V(\phi)}\biggm\vert_{\phi=0}
\end{equation}
Similarly, any correlator is given by
\begin{equation}
<\tr(\bar\psi\psi)^{p_1}\tr(\bar\psi\psi)^{p_2}\cdots>~=
\frac{\det^N\left(\frac{\partial}{\partial\phi}\right)\tr\phi^{p_1}
\tr\phi^{p_2}\cdots\e^{N^2\tr V(\phi)}}{\det^N
\left(\frac{\partial}{\partial\phi}
\right)\e^{N^2\tr V(\phi)}}\biggm\vert_{\phi=0}
\end{equation}
In spite of this good convergence of the partition function, it has
been shown that in the infinite $N$ limit,
the model has third order phase transitions \cite{mz,akm}. In this
Letter, we shall examine the third order transitions in more detail
and also argue that there may be a first order phase transition.  We show that
in a simple model with potential $V(\ps2)=\bar\psi\psi+\frac{g}{3}(\bar\psi
\psi)^3$ there are two critical
points, one of them at the third order phase transition at
$g=g_c\equiv 2/27$ which was studied in \cite{mz,akm} and the other at
zero coupling $g=0$.  In the region $g\in[0,g_c]$, multi-branch
one-cut solutions exist as well as a multi-cut solution of the matrix
model whereas when $g\notin[0,g_c]$ a unique one-cut solution with real
free energy exists.

At $g=g_c$ the unique one-cut solution for $g>g_c$ connects
continuously with the one of the 3 one-cut solutions which has the minimal
free energy in $0<g<g_c$.  On the other hand, this solution is not
continuous with the unique one-cut solution for $g<0$. The one-cut
solution for $g<0$ connects to one which does not have minimal free
energy.  Ordinarily, the infinite energy barrier (the height of the
barrier is of order $N^2$) between the metastable and stable one-cut
solutions prevents tunneling and also a phase transition from
occuring.  However, if we restrict attention to one-cut solutions, and
follow them over the range of $g$, we must encounter a discontinuity
of the free energy somewhere, i.e. a first order phase transition.
The choice of stable solution (the one-cut solution with minimum free
energy) in the regime $g\in[0,g_c]$ results in a first order phase
transition at $g=0$. With this stability requirement
perturbation theory near the Gaussian point $g=0$ does
not correctly reflect the properties of the theory when $g$ is small
and negative. The critical point $g_c$ corresponds to the usual $m=2$
multi-critical point with $\gamma_{\rm str}=-1/2$. In the following, we shall
also explicitly construct one-cut solutions for generic symmetric
potentials and thus show how higher order multi-critical points can be
realized in these models.

First, we shall discuss some general properties of the adjoint fermion
matrix model (\ref{part}). It possesses a continuous symmetry
$\psi\rightarrow U\psi V^{-1}$, $\bar\psi\rightarrow V\bar\psi U^{-1}$
with $\{U,V\}\in GL(N,\IC)
\otimes GL(N,\IC)$. In spite of this large degree of symmetry, it is not
possible to diagonalize a matrix with anticommuting entries. Thus,
unlike the more familiar Hermitian one-matrix models, the model
(\ref{part}) cannot be written as a statistical theory of eigenvalues.
Nevertheless, in the large-$N$ limit it shares many of the properties
of such a theory \cite{bipz,biz}.  Furthermore the large degree
of symmetry restricts the observables to those which are essentially
invariant functions of $\bar\psi\psi$. The chiral transformation
$\psi\to\bar\psi$, $\bar\psi\to-\psi$,
$\tr(\bar\psi\psi)^k\to(-1)^{k+1}\tr(\bar\psi\psi)^k$ is a symmetry
when the potential is an {\it odd} polynomial.  Furthermore, in that
case, all {\it even} moments vanish, $<\tr (\bar\psi\psi)^{2k}>~=0$.
This is the analog of a symmetric potential for a Hermitian matrix
model \cite{bipz}.

In the large-$N$ limit, correlators of the matrix model factorize
\begin{equation}
<{\tr}f(\ps2){\tr}g(\ps2)>~=~<\tr f(\ps2)>~<\tr g(\ps2)>+{\cal O}(1/N^2)
\end{equation}
This factorization property follows from the existence of a finite
large-$N$ limit for the correlators $<\tr(\bar\psi\psi)^k>$ for
arbitrary polynomial potential $V(\ps2)=\sum_k\frac{g_k}{k}(\bar\psi\psi)^k$,
since then the connected correlators are given by
\begin{equation}
<\tr(\bar\psi\psi)^p\tr(\bar\psi\psi)^k>_{\rm conn} =
\frac{1}{N^2}p\frac{\partial}{\partial g_p} <\tr(\bar\psi\psi)^k>~\sim~
\frac{1}{N^2}
\end{equation}
Factorization and symmetry imply that the large-$N$ limit of the model
is completely characterized by the set of correlators $<\tr(\bar\psi\psi)^k>$.
Since the signs of the actions in (\ref{part}) and (\ref{penner})
are opposite (from the different boundary conditions), it also follows that
the connected correlators of the fermionic matrix model alternate in sign
relative to those of the generalized Hermitian Penner model (\ref{penner}).
It is this property that leads to an alternating series for the large-$N$ genus
expansion of the fermionic matrix model (\ref{part}).

When $N$ is finite, nilpotency of the components of $\psi$ and
$\bar\psi$ implies that the moments $<\tr(\bar\psi\psi)^k>$ are
non-zero only for $k\leq N^2$ and are therefore finite in number.  In
this case, the moment generating function
\begin{equation}
\omega(z)=~<\tr\frac{1}{z-\bar\psi\psi}>~
=\sum_{k=0}^{N^2}<\tr(\bar\psi\psi)^k>\frac{1}{z^{k+1}}
\label{gen}
\end{equation}
has singularities only at the origin in the complex $z$-plane.
The finite set of moments can always be obtained from a (not unique)
distribution function $\rho$ with support in the complex plane
\begin{equation}
<\tr(\bar\psi\psi)^k>~=\int d\alpha~\rho(\alpha)\alpha^k~~~{\rm with}~~~
\int d\alpha~\rho(\alpha)=1
\label{mom}
\end{equation}
The support of $\rho$ can be deduced from the position of the
singularities of $\omega$ in (\ref{gen}).  When $N$ and therefore the
number of moments is finite the support of $\rho$ is concentrated near
the origin
\begin{equation}
\rho(\alpha)=\sum_{k=0}^{N^2}\frac{1}{k!}<\tr(\bar\psi\psi)^k>\left(-
\frac {\partial}{\partial\alpha}\right)^k \delta(\alpha)
\end{equation}
In the large-$N$ limit, the spectral function $\rho(\alpha)$ can be a
function with support on some contour in the complex plane.  The
distribution function $\rho$ is the analog in the fermionic matrix
model of the density of eigenvalues in Hermitian one-matrix models
as the quantity which specifies the solution of the model in the
infinite $N$ limit \cite{bipz}.

The generating functions for the connected correlators are
\begin{equation}
\omega_n(z_1,\ldots,z_n)=~<\tr\frac{1}{z_1-\ps2}\cdots
\tr\frac{1}{z_n-\bar\psi\psi}>_{\rm conn}
\end{equation}
In particular, the loop correlator
\begin{equation}
\omega_1(z)\equiv\omega(z)=\int d\alpha~\frac{\rho(\alpha)}{z-\alpha}
\label{omdef}
\end{equation}
is analytic in $z$ away from the support of $\rho$ in the complex
plane.  The distribution function can be determined by computing the
discontinuity of $\omega(z)$ across its support, where
\beq
\omega(\lambda\pm \epsilon_\perp)
=\pvint d\alpha~\frac{\rho(\alpha)}{\lambda-\alpha}
\mp \frac{\epsilon_\perp}{\vert\epsilon_\perp\vert}
\pi\rho(\lambda)~~~{\rm for}~~~\lambda\in{\rm supp}~\rho
\label{disc}
\eeq
where $\epsilon_\perp(\lambda)$ is a complex number with infinitesimal
amplitude and direction perpendicular to the integration contour at
the point $\lambda$.  The Schwinger-Dyson equation for the loop correlator
$\omega(z)$ can be derived from the invariance of the partition
function (\ref{part}) under arbitrary changes of variables. It can be
cast in the form \cite{mz,akm}
\beq
\int d\psi~d\bar\psi~\frac{\partial}{\partial\psi_{ij}}\left[\left(\psi\frac{
1}{z-\ps2}\right)_{k\ell}\e^{N^2\tr V(\ps2)}\right]=0
\label{loop1}
\eeq
In contrast to Hermitian matrix models \cite{kaz}, the identity (\ref{loop1})
is exact for fermionic matrices. It leads to
\beq
-z \omega(z)^2+\left(2-z V'(z)\right)\omega(z)+V'(z)+P(z)=z\omega_2(z,z)
\label{loop}
\eeq
where $P(z)$ is a polynomial of degree $K-2$
\beq
P(z)=\sum_{k=2}^K g_k\sum_{p=0}^{k-2}<\tr(\bar\psi\psi)^{k-1-p}>z^p
\label{polynomial}
\eeq

Factorization implies that the connected correlators are all
suppressed by factors of $1/N^2$ and the term on the right-hand side
of the loop equation (\ref{loop}) vanishes in the large-$N$ limit.
Then the loop equation has solution
\begin{equation}
\omega(z)=\frac{1}{z}-\frac{V'(z)}{2}+\frac{1}{z}\sqrt{ 1+
\left(\frac{zV'(z)}{2}\right)^2+zP(z)}
\label{omcut}
\end{equation}
where the sign of the square root is chosen to yield the correct asymptotic
behavior $\omega(z)\to1/z$ at $|z|\to\infty$. The branches of the square root
must be placed so that it is negative near the origin in order to cancel the
pole at $z=0$. If the potential is a polynomial of order $K$, the solution
will in general possess a square root singularity with $K$ branch cuts
and the spectral density $\rho$ will have $K$ contours in its support.

The simplest solution of the model is the one-cut solution which assumes that
the singularities of $\omega(z)$ consist of only a single square root branch
cut, so that the distribution $\rho$ has support only on one arc in the
complex plane with endpoints at some complex values $a_1$ and $a_2$. This
solution for $\omega(z)$ can be represented in the form \cite{mz,akm}
\beq
\omega(z)=\oint_{\cal C}\frac{dw}{4\pi i}\frac{V'(w)-2/w}{z-w}
\sqrt{\frac{(z-a_1)(z-a_2)}{(w-a_1)(w-a_2)}}
\label{omcont}
\eeq
where the closed contour $\cal C$ encloses the support of the spectral function
but not the point $w=z$. The endpoints of the cut can then be found by
imposing the asymptotic boundary condition $\omega(z)\to1/z$ at $|z|\to
\infty$ on the solution (\ref{omcont}), which leads to the two equations
\beq
\oint_{\cal C}\frac{dw}{2\pi i}\frac{V'(w)-2/w}{\sqrt{(w-a_1)(w-a_2)}}=0~~~,
{}~~~\oint_{\cal C}\frac{dw}{2\pi i}\frac{wV'(w)-2}{\sqrt{(w-a_1)(w-a_2)}}=2
\label{1cut}
\eeq

To determine the precise location of the support contour of $\rho$ in the
complex plane, we first use the observation of \cite{mz} that the large-$N$
equation (\ref{loop}) for the loop correlator is identical to the loop
equation for the generalized Penner model (\ref{penner}). In the large-$N$
limit, the spectral density therefore obeys the saddle-point equation
\cite{bipz}
\beq
\frac{2/\lambda-V'(\lambda)}{2}=\pvint d\alpha~\frac{\rho(\alpha)}{\lambda-
\alpha}~~~~,~~~~\lambda\in~{\rm supp}~\rho
\label{saddle}
\eeq
Note that this equation can be obtained from the discontinuity (\ref{disc})
of the loop correlator (\ref{omcut}), and it also follows from the
minimization condition for the free energy
\beq
F=\lim_{N\to\infty}\frac{1}{N^2}\log Z_P=\int d\alpha~\rho(\alpha)\left(
V(\alpha)-2\log\alpha\right)+\int\!\!\pvint d\alpha~d\lambda~\rho(\alpha)
\rho(\lambda)\log(\alpha-\lambda)
\label{free}
\eeq
with respect to the distribution function $\rho$. Note the change in sign of
the fermionic free energy relative to the Hermitian case. The double
integral in (\ref{free}) is evaluated by integrating up the saddle-point
equation (\ref{saddle}). This introduces a logarithmic divergence at
$\lambda=0$ arising from the Penner potential in (\ref{penner}) which we
remove by subtracting from (\ref{free}) the Gaussian free energy $F_G$
defined by setting $g_k=0$ for $k>1$ in (\ref{free})
\beq
F-F_G=\frac{1}{2}\int d\alpha~\rho(\alpha)\left(V(\alpha)-2\log\alpha\right)
+\pvint d\alpha~\rho(\alpha)\log\alpha
\label{freeg}
\eeq
where we have ignored terms independent of $g_k$, $k>1$. The support contour
of $\rho$ can be determined from the David primitive function
\cite{david}
\beq
G(w)=\int_{a_1}^wdz~\left(\frac{2}{z}-V'(z)-2\omega(z)\right)
\label{david}
\eeq
The support of $\rho$ is an arc connecting $a_1$ to $a_2$ in the complex
plane along which $G(w)$ is purely imaginary and which can be embedded in a
region where ${\rm Re}~G(w)<0$ \cite{david}.

For illustration, we shall consider the cubic potential
\beq
V(z)=tz+\frac{g}{3}z^3
\label{potc}
\eeq
for which
\beq
\omega(z)=\frac{1}{z}-\frac{t}{2}-\frac{g z^2}{2}+\frac{1}{2 z}
\sqrt{g^2 z^6+2tg z^4+(t^2+4g\xi ) z^2+4}
\label{ocubic}
\eeq
where $\xi $ is the as yet unknown correlator $\xi =~<\tr\ps2>$. In this
case the vanishing of all even moments, $\int d\alpha~\rho(\alpha)
\alpha^{2k}=0$, implies that the endpoints of the support contour of the
continuous function $\rho$ lie in the complex plane and are symmetric on
reflection through the origin. Furthermore, an application of Wick's
theorem shows that the series (\ref{gen}) in the odd moments is alternating.

Generically the square root in $\omega(z)$ has three branch cuts, so
that in the general case the distribution $\rho$ will have three
disjoint and symmetric (about the origin) support contours.
The one-cut solution for (\ref{ocubic}) takes the form
\beq
\omega(z)=\frac{1}{z}-\frac{t}{2}-\frac{g z^2}{2}+\frac{gz^2\pm b}{2z}
\sqrt{ z^2+4/b^2}
\label{om1cut}
\eeq
where comparing the polynomial coefficients in (\ref{om1cut}) with
those of (\ref{ocubic}) shows that the parameter $b$ and the
correlator $\xi $ are determined by the two equations
\begin{equation}
\pm b^3-tb^2+2g=0
\label{b}
\end{equation}
\begin{equation}
b^3-(t^2+4g\xi )b\pm8g=0
\label{C}
\end{equation}
The sign ambiguity here can be eliminated by requiring that at $g=0$ the
correct
Gaussian value $b(g=0,t)=t$ for $b$ \cite{mz} be attainable. This is the
boundary condition that is relevant for an interpretation of this matrix model
as a discretized random surface theory, i.e. for a consistent perturbative
expansion of the model in the coupling constant $g$. It means that
we take the positive sign in the above equations. The choice of negative
sign yields solutions with boundary conditions at $g=0$ appropriate to
generalized Penner models \cite{akm,pen}. The equation (\ref{b}) and this sign
ambiguity also follow from the contour integrals (\ref{1cut}).

We assume henceforth that $t$ is a positive constant. The 3
solutions of (\ref{b}) are
\begin{equation}
b_0(x,t)=\frac{t}{3}\left(\beta^{1/3}(x)+\beta^{-1/3}(x)+1\right)
\label{bb}
\end{equation}
\beq
b_\pm(x,t)=\frac{t-b_0(x,t)}{2}\pm\frac{i\sqrt{3}t}{6}\left(\beta^{1/3}(x)-
\beta^{-1/3}(x)\right)
\label{bim}
\eeq
where
\beq
\beta(x)=2x-1+2\sqrt{x(x-1)}
\label{beta}
\eeq
and we have introduced the scaling
parameter $x=1-\frac{27g}{2t^3}$. When $x\leq0$ ($g\geq g_c\equiv\frac{
2t^3}{27}$) or $x\geq1$ ($g\leq0$), $\beta(x)$ is a monotone real-valued
function with $\beta(x)\geq1$ for $x\geq1$ and $\beta(x)\leq-1$ for $x\leq0$.
In the region $0<x<1$ ($0<g<g_c$), $\beta(x)$ is a complex-valued function
with unit modulus. The function (\ref{bb}) is always real-valued and the
region $0<x<1$ is the region wherein all 3 roots (\ref{bb}), (\ref{bim})
of the cubic equation (\ref{b}) are real. These 3 roots can all be obtained
from (\ref{bb}) by choosing the 3 inequivalent cube roots of $\beta(x)$.
For $x\notin(0,1)$ the solutions (\ref{bim}) are complex. For the
fermionic matrix model, where the distribution function $\rho$ can be
complex-valued, there is no immediate reason to disregard complex-valued
endpoints for the support of $\rho$. However, the free energy (\ref{freeg})
for the cubic potential (\ref{potc}) up to terms independent of $b$ and
$g$ is
\beq
F(x,t)-F_G(t)=\frac{t(3b(x,t)-t)}{6b^2(x,t)}
\label{freec}
\eeq
where we have used the spectral density determined by (\ref{disc}) and
(\ref{om1cut})
\beq
\rho(\alpha)=\frac{1}{2\pi i}\left(b+g\alpha^2\right)\sqrt{1+\frac{4}
{b^2\alpha^2}}~~~~;~~~~\alpha\in{\cal C}_b
\label{rho}
\eeq
where ${\cal C}_b=~{\rm supp}~\rho$ and $b(g,t)$ are given by (\ref{bb}) and
(\ref{bim}). It is immediately seen that the free energy (\ref{freec}) is
complex-valued for the values (\ref{bim}) of $b(x,t)$ for $x\notin(0,1)$.
Such a free energy leads to an unstable state and we therefore consider
only the real-valued solutions to (\ref{b}).

The support contour ${\cal C}_b$ on which (\ref{rho}) is defined is found
from the David function (\ref{david}) which is
\bea
G(z)=-\sqrt{b^2z^2+4}-~{\rm sgn}~(b)\log\left(\frac{\sqrt{b^2z^2+4}-2}
{\sqrt{b^2z^2+4}+2}\right)-\frac{g}{3|b|^3}\left(b^2z^2+4\right)^{3/2}
\nonumber\\ +\frac{g}{b^2}\log\left[\frac{i}{2}\left(\sqrt{b^2z^2+4}+|b|z
\right)\right]-i\pi~{\rm sgn}~b
\label{davidc}
\eea
where the branch of the square roots in (\ref{davidc}) is taken to be the
straight line joining the points $\pm2i/|b|$. A careful examination of
the equation ${\rm Re}~G(z)=0$ and of the region
where ${\rm Re}~G(z)<0$ shows that the contour ${\cal C}_b$ cannot cross
the imaginary axis for $|~{\rm Im}~z|>2/|b|$ and that it crosses the real
axis at some non-zero values of order $\pm1/|b|$. The regions ${\rm Re}~
G(z)<0$ are to the right of these crossing points (but note that ${\rm Re}~
G(z)$ changes sign across ${\cal C}_b$). Thus the contour ${\cal C}_b$ in
(\ref{rho}) can be taken as the counterclockwise oriented half-circle of
radius $2/|b|$ in the first and fourth quadrants of the complex
$\alpha$-plane. It is easy to verify that with this definition of $\rho$
the equations (\ref{mom}) and (\ref{omdef}) are satisfied, as is (\ref{C})
from evaluating the correlator $\xi=\int d\alpha~\rho(\alpha)\alpha$ with
this distribution function.

There are 2 critical points in this large-$N$ matrix model, at $g=0$ and
$g=g_c$, which separate 3 phases determined by the analytic structure of the
function (\ref{beta}), i.e. the one-cut solution is a non-analytic function
of $x$ about $x=0$ and $x=1$ where it acquires a square root branch cut.
For $x\geq1$ the solution
\begin{equation}
b_0(x,t)=\frac{t}{3}\left[1+\left(2x-1+2\sqrt{x(x-1)}\right)^{1/3}+
\left(2x-1+2\sqrt{x(x-1)}\right)^{-1/3}\right]~,~x\geq1
\label{x>1}
\end{equation}
of (\ref{b}) satisfies the Gaussian boundary condition $b_0(x=1,t)=t$. When
$x\leq 0$ the real solution for $b$ is
\begin{equation}
b_0(x,t)=\frac{t}{3}\left[1-\left|2x-1+2\sqrt{x(x-1)}\right|^{1/3}-
\left|2x-1+\sqrt{x(x-1)}\right|^{-1/3}\right]~,~x \leq 0
\label{x<0}
\end{equation}

As $x$ is varied between 0 and 1, $\beta(x)$ has modulus one and phase which
varies from $\pi$ to $0$: $\beta(x)=\e^{i\phi(x)}$ where
\begin{equation}
\phi(x)=\arctan\left(\frac{2\sqrt{x(1-x)}}{2x-1}\right)\in[0,\pi]
\label{phi}
\end{equation}
The arctangent function in (\ref{phi}) is well-defined only up to an
integral multiple of $2\pi$, and the three real solutions for $b$ are
\begin{equation}
b(x,t)=\frac{t}{3}\left[1+2\cos\left\{\frac{1}{3}\arctan\left(\frac
{2\sqrt{x(1-x)}}{2x-1}\right)+\frac{2n\pi}{3}
\right\}\right]~~~;~~~0<x<1~,~n=0,1,2
\label{0<x<1}
\end{equation}
The branch which matches (\ref{x<0}) is the one with $n=1$, whereas
the branch which matches (\ref{x>1}) is the one with $n=0$. The
branch with $n=2$ does not connect with either solution. Any of these 3
branches can be used to define the one-cut solution (\ref{om1cut}). The
free energy (\ref{freec}) is positive for all $x\in(0,1)$ for the $n=0$
branch, negative for all $x\in(0,1)$ for the $n=1$ branch, and for the
$n=2$ branch it is positive for $0<x<\frac{1}{2}$ and flips sign for the
rest of the interval at $x=\frac{1}{2}$. The $n=1$ branch in (\ref{0<x<1})
is the ground state solution in the region $0<x<1$.

The free energy associated with this stable one-cut solution is
discontinuous across $g=0$, and thus with this choice of branch in the regime
$0<x<1$ the Gaussian point of this matrix model is a critical point of a
first order phase transition. The other one-cut solution which is a
perturbation of the Gaussian solution is metastable but can still be
thought of as a valid solution of the model since the barrier is infinite
at $N=\infty$. This is similar to the situation in the Hermitian one-matrix
model with symmetric polynomial potential of degree 6 \cite{jurk}. There a
phase transition occurs due to an infinite volume effect, as opposed to a
large-$N$ effect where the only possibilities could be second or third
order phase transitions. There is also the possibility that the loop
correlator (\ref{ocubic}) evolves into a three-cut phase at $g=0$,
corresponding to a third order phase transition at $g=0$ (see below), but
there is no immediate indication of this since in the fermionic case the
spectral measure $\rho(\alpha)~d\alpha$ need not be positive. This possibility
is also suggested by the exact form (\ref{ocubic}) of the loop correlator:
Although the one-cut ansatz (\ref{om1cut}) is insensitive to a change in
sign of $g$, the analytic properties of (\ref{ocubic}) might be affected
by the passage through $g=0$.

The existence of 3 phases in this matrix model and the possibility of
a first order phase transition at the Gaussian point $g=0$ are completely
unlike what occurs in the conventional polynomial Hermitian matrix models
\cite{bk,kaz} or in Penner models \cite{akm,cdl}. The unusual phase
transition at $g=0$ would imply that the consistent perturbative expansion
of the theory in $g$ has a preferred direction through values of $g$ with
a definite sign (corresponding to $-~{\rm sgn}~t$). This fact is important
for the interpretation of the fermionic one-matrix model as a
statistical theory of discretized random surfaces. Notice, however, that
the free energy (\ref{freec}) with the choice of stable branch for
$x\in(0,1)$ is continuous across the point $g=g_c$ which coincides with the
critical point found in \cite{akm}.

The scaling behaviour of the matrix model in the vicinity of its
critical points is determined by the string susceptibility
\beq
\chi(g,t)=-\frac{1}{N^2}\frac{\partial^2\log Z}{\partial g^2}=-\frac{1}{3}
\frac{\partial}{\partial g}<\tr(\ps2)^3>
\label{chi}
\eeq
The critical exponent $\gamma_{\rm str}^{(i)}$ at each critical point
$g_c^{(i)}$ is defined by the leading non-analytic behaviour of (\ref{chi})
\cite{bk}
\beq
\chi(g,t)\sim_S(g-g_c^{(i)})^{-\gamma_{\rm str}^{(i)}}~~~{\rm as}~~~g
\to g_c^{(i)}
\end{equation}
where $\sim_S$ denotes the most singular part of the function in a
neighbourhood of the critical point. In terms of the scaling variable
$x$, the susceptibility (\ref{chi}) is
\beq
\chi(x,t)=\frac{972}{t^6(\beta^{2/3}+\beta^{1/3}+1)^6}\left[1-8x+8x^2+4\sqrt{
\frac{x}{x-1}}\left(1-3x+2x^2\right)\right]
\label{chi1}
\eeq
{}From (\ref{chi1}) we find that the leading singular parts of the
susceptibility
near each of the two critical points $g=g_c$ and $g=0$ are respectively
\beq
\chi(x,t)\sim_S-\frac{15552}{t^6}\sqrt{x}~~~{\rm as}~~~x\to0
\label{sc1}
\eeq
\beq
\chi(x,t)\sim_S\frac{11648}{3t^6}\sqrt{x-1}~~~{\rm as}~~~x\to1
\label{sc2}
\eeq
Both critical points therefore have string constant $\gamma_{\rm str}=-
1/2$ which coincides with those of the usual genus zero $m=2$ quantum
gravity models \cite{kaz}.

In particular, this shows that with the choice of stable branch for
$x\in(0,1)$ the phase transition at the non-zero critical coupling $g=g_c$
is of third order. Indeed, the critical point $g=g_c$
enjoys all of the properties of a conventional $m=2$ multi-critical
point \cite{kaz}. The information relevant to the topological genus
expansion in string theory is contained in this non-zero third order
critical coupling. As argued in \cite{akm}, this $m=2$ multi-critical
point results in a genus expansion which is an alternating series but
otherwise coincides with the usual Painlev\'e expansion \cite{bk}.

Thus the 2 critical points of the fermionic one-matrix model, which arise
as those points in parameter space where the cubic equation (\ref{b}) which
determines the one-cut solution acquires a double real root, provide the
necessary information for the perturbative and topological expansions of the
theory. In the 2 phases outside of the region $0<x<1$ the one-cut solution
is unique, while in the phase $x\in(0,1)$ a three-cut solution can in addition
exist. The scaling behaviours (\ref{sc1}) and (\ref{sc2}) indicate that the
2 transitions into the multi-cut phase would both be of third order, while
the transitions into the stable one-cut phase are of first and third order.
The existence of a single-cut or multi-cut phase in $0<x<1$ is determined
by which one of these 2 possibilities is in fact the vacuum state.
The phase structure of this model is reminiscent of the triple
point of a liquid-vapour-solid system, where there is a classical
forbidden region in the liquid phase for the isotherms to pass through. The
system can however supercool from the vapour to solid phase by tunneling
through this region which corresponds to a first order phase transition,
while the other phase transitions are of third order. In our case this
``forbidden" region is $0<x<1$ (as this is where the function $\beta(x)$
becomes
non-analytic) and it has width $g_c=2t^3/27$. The triple point where all
3 phases coexist is then obtained by letting $t\to0$.

We conclude by discussing the critical behaviour associated
with higher order potentials. For simplicity we consider the
chirally symmetric case where the potential (\ref{pot}) is a generic
odd polynomial, i.e. $g_{2k}=0$ for all $k$ in (\ref{pot}), with
$K=\deg V>3$ an odd integer. The solution for the loop correlator is
then
\beq
\omega(z)=\frac{1}{z}-\sum_{k=1}^{\frac{K+1}{2}}\frac{g_{2k-1}}{2}z^{2(k-1)}
+\frac{1}{2z}\left(4+\sum_{k,m=1}^{\frac{K+1}{2}}g_{2k-1}g_{2m-1}z^{2(k+m-1)}
+\sum_{k,m=1}^{m+k\leq\frac{K+1}{2}}g_{2(k+m)-1}\xi_{2m}z^{2k}\right)^{1/2}
\label{omK}
\eeq
where $\xi_{2m}$ are the as yet unknown moments
$\xi_{2m}=~<\tr(\ps2)^{2m-1}>$. The one-cut solution for
(\ref{omK}) takes the form
\beq
\omega(z)=\frac{1}{z}-\sum_{k=1}^{\frac{K+1}{2}}\frac{g_{2k-1}}{2}z^{2(k-1)}+
\frac{1}{2z}\left(g_Kz^{K-1}+\sum_{k=1}^{\frac{K-3}{2}}a_{2k}z^{2k}+b\right)
\sqrt{z^2+4/b^2}
\label{K1cut}
\eeq
where we have fixed the sign in front of the parameter $b$ by the same
convention as before. The one-cut ansatz (\ref{K1cut}) along with the
general solution (\ref{omK}) together involve $K-1$ unknown parameters
-- $b$, the $(K-3)/2$ polynomial coefficients $a_{2k}$, and the
$(K-1)/2$ correlators $\xi_{2k}$.

These parameters can be found by comparing the various polynomial coefficients
of (\ref{omK}) with those of (\ref{K1cut}), which leads to the set of equations
\beq
b^3+8a_2-b\left(g_1^2+\sum_{k=1}^{\frac{K-1}{2}}g_{2k+1}\xi_{2k}\right)=0~~~~
,~~~~2b^3+8ba_4+4a_2^2-b^2\left(2g_1g_3+\sum_{k=1}^{\frac{K-3}{2}}g_{2k+3}
\xi_{2k}\right)=0
\label{parstart}
\eeq
\beq
2bg_Ka_{K-5}+8a_{K-3}-b\sum_{k=1}^{K-2}g_{2k-1}g_{2K-2k-3}=0~~~~,~~~~2b^2g_K
a_{K-3}+4g_K^2-b^2\sum_{k=1}^{K-1}g_{2k-1}g_{2K-2k-1}=0
\eeq
\beq
2b^2a_{K-3}+8g_K-b\left(g_K\xi+\sum_{k=1}^{\frac{K-1}{2}}g_{2k-1}g_{K-2k}
\right)=0~~~~,~~~~2b^3g_K+8bg_Ka_2-b^2\sum_{k=1}^{\frac{K+1}{2}}g_{2k-1}
g_{K-2k+2}=0
\eeq
When $K>7$ we have in addition the sets of equations
\bea
2b^3a_{2(m-1)}+8ba_{2m}+4a_2a_{2(m-1)}+\sum_{k=1}^{m-2}\left(b^2a_{2k}a_{2
(m-k-1)}+4a_{2k}a_{2(m-k)}\right)\nonumber\\-b^2\left(\sum_{k=1}^mg_{2k-1}
g_{2(m-k)+1}+\sum_{k=1}^{\frac{K+1-2m}{2}}g_{2(m+k)-1}\xi_{2k}\right)=0
\eea
for $3\leq m\leq\frac{K-3}{2}$, and
\bea
2b^2g_Ka_{2m-K-1}+8ba_{2m-K+1}+b^2\sum_{k=1}^{m-2}a_{2k}a_{2(m-k-1)}\nonumber
\\+4\sum_{k=1}^{m-1}a_{2k}a_{2(m-k)}-b^2\sum_{k=1}^mg_{2k-1}g_{2(m-k)+1}=0
\label{parend}
\eea
for $\frac{K+3}{2}\leq m\leq K-3$.

(\ref{parstart})--(\ref{parend}) yield a complete set of equations for
the $K-1$ unknown coefficients of the one-cut solution (\ref{K1cut})
in terms of the coupling constants of the potential (\ref{pot}). The
parameter $b$ can alternatively be found from the contour
integrals (\ref{1cut}) which lead to a $K$-th order equation for $b$
\beq
b^K+\sum_{k=1}^{\frac{K+1}{2}}\frac{(-1)^k2^{k-1}(2k-3)!!}{(k-1)!}g_{2k-1}
b^{K-2k+1}=0
\label{bK}
\eeq
Since $K$ is odd this equation always has a real solution, and as
before the one-cut solution can always be constructed. The spectral
density is
\beq
\rho(\alpha)=\frac{1}{2\pi i}\left(g_K\alpha^{K-1}+\sum_{k=1}^{\frac{K-3}{2}}
a_{2k}\alpha^{2k}\right)\sqrt{1+\frac{4}{b^2\alpha^2}}~~~;~~~\alpha\in{\cal
C}_b
\label{distrK}
\eeq
The first $(K-1)/2$ moments of this distribution are given by the
solutions to (\ref{parstart})--(\ref{parend}).

In general (\ref{bK}) will acquire multiple real roots at some
coupling $g_{2k-1}^c$ which will be a critical point of a third order
phase transition with string constant $\gamma_{\rm str}=-1/2$. We
expect that the phase with multiple real roots will be bounded by
other coupling constant values so that the model will contain several
critical points corresponding possibly to different order phase
transitions. These critical points might provide information about this matrix
model other than that relevant to the genus expansion. Notice that since
the potential now depends on more parameters, we can adjust them in such a
way that $m-1$ regular zeroes of (\ref{distrK}) coalesce with a cut end-point
$\pm2i/|b|$ at criticality for the same critical coupling $g_{2k-1}^c$.
$g_{2k-1}^c$ will then be an $m$-th order multi-critical point with
susceptibility exponent $\gamma_{\rm str}=-1/m$ \cite{kaz}. In this way the
construction above shows explicitly how more general two-dimensional quantum
gravity theories \cite{bk,kaz} may be obtained from continuum
limits of fermionic matrix models.

\vskip 1cm

The authors gratefully thank Yu. Makeenko for helpful discussions and
comments on the manuscript. We also thank L. Paniak for helpful discussions.


\clearpage\newpage

\begin{thebibliography}{10}

\baselineskip=12pt

\bibitem{tH} G. 'tHooft, Nucl. Phys. {\bf B72} (1974), 461.

\bibitem{bipz} E. Br\'ezin, C. Itzykson, G. Parisi and J.-B. Zuber, Commun.
Math. Phys. {\bf 59} (1978), 35.

\bibitem{biz} D. Bessis, C. Itzykson and J.-B. Zuber, Adv. Appl. Math.
{\bf 1} (1980), 109.

\bibitem{adf} J. Ambj\o rn, B. Durhuus and J. Fr\"ohlich, Nucl. Phys. {\bf
B257} [FS14] (1985), 433; F. David, Nucl. Phys. {\bf B257} [FS14]
(1985), 45; V. A. Kazakov, Phys. Lett. {\bf B150} (1985), 282; V. A.
Kazakov, I. K. Kostov and A. A. Migdal, Phys. Lett. {\bf B157} (1985),
295.

\bibitem{bk} E. Br\'ezin and V. A. Kazakov, Phys. Lett. {\bf B236} (1990),
144; M. R. Douglas and S. H. Shenker, Nucl. Phys. {\bf B335} (1990),
635; D. J. Gross and A. A. Migdal, Phys. Rev. Lett. {\bf 64} (1990),
127.

\bibitem{gw} D. J. Gross and E. Witten, Phys. Rev. {\bf D21} (1980), 446.

\bibitem{di} A. Drouffe and J.-B. Zuber, Phys. Rep. {\bf 102} (1983), 1.

\bibitem{km} V. A. Kazakov and A. A. Migdal, Nucl. Phys. {\bf B397} (1993),
214.

\bibitem{mz} Yu. Makeenko and K. Zarembo, Nucl. Phys. {\bf B422} (1994), 237.

\bibitem{akm} J. Ambj\o rn, C. F. Kristjansen and Yu. Makeenko:
Generalized Penner Models to all Genera, Niels Bohr Institute preprint
NBI-HE-94-14 (1994).

\bibitem{khm} S. Khokhlachev and Yu. Makeenko, Mod. Phys. Lett. {\bf A7}
(1992), 3653; A. A. Migdal, Mod. Phys. Lett. {\bf A8} (1993), 359.

\bibitem{kaz} V. A. Kazakov, Mod. Phys. Lett. {\bf A4} (1989), 2125.

\bibitem{pen} R. C. Penner, Commun. Math. Phys. {\bf 113} (1987), 299; J. Diff.
Geom. {\bf 27} (1988), 35.

\bibitem{david} F. David, Nucl. Phys. {\bf B348} (1991), 507.

\bibitem{jurk} J. Jurkiewicz, Phys. Lett. {\bf B245} (1990), 178.

\bibitem{cdl} S. Chaudhuri, H. M. Dykstra and J. D. Lykken, Mod. Phys. Lett.
{\bf A6} (1991), 1665.

\end{thebibliography}
\end{document}